\documentclass[preprintnumbers,nofootinbib,a4paper,superscriptaddress]{revtex4-1}

\usepackage{amsmath,latexsym,amssymb,amsfonts}
\usepackage{amsthm}

\usepackage{hyperref}

\newtheorem{theorem}{Theorem}

\begin{document}

\title{Twinor Theory in the Newton--Hooke Limit\texorpdfstring{\\}{: }
A Nonrelativistic Schrödinger Equation for Arbitrary Spin}
\author{Chen-Hsu Chien}
\email[E-mail: ]{chien@math.muni.cz}
\affiliation{Department of Mathematics and Statistics, Masaryk University}
\date{\today}

\begin{abstract}
The Newton--Hooke limit of the massive twinor field equations is derived at fixed particle mass and fixed cosmological acceleration per unit displacement. On a static torsion-free background, removal of the rest energy separates the leading spin amplitude from a finite sequence of auxiliary spinors. For every positive integer rank $2s$, a regular limit of exact solutions satisfies a first-order hierarchy. Its lowest two equations give a necessary Schr\"odinger equation for the leading amplitude, with a curvature term from the ordered spatial derivatives and a source determined by two projections of the relativistic source. The higher equations impose additional conditions. On the homogeneous Newton--Hooke background with trivial internal connection and nonzero cosmological acceleration parameter, vanishing of all limiting contracted sources forces the leading amplitude to vanish for $s\geq\tfrac32$. A separate reduction of the scalar wave equation gives the spin-zero case.

\end{abstract}

\maketitle

\section{Introduction}

In the Newton--Hooke limit, a freely falling particle retains a finite acceleration proportional to its displacement from the origin. The corresponding potential energy is quadratic in position. Two questions arise. How does this potential enter the field equation of a massive particle of spin $s$? Which constraints on the relativistic field survive the limit? The latter question arises because a symmetric multispinor satisfies a first-order equation in each of its spinor slots.

To specify the limit, let $c$ be the speed of light and $l$ the cosmological curvature scale, defined by $\Lambda=-3/l^2$ for cosmological constant $\Lambda$. Both the de Sitter and anti-de Sitter real forms are included by allowing $l$ to be complex. As $c\to\infty$, the time scale $\kappa=l/c$ is held fixed. The signed quantity $\Upsilon^2$, with dimensions of inverse time squared, then stays finite. In Cartesian spatial coordinates $x$ and Newtonian time $t$ the scaling, potential per unit mass and equation of free fall are
\begin{equation*}
\Upsilon^2=\frac{c^2\Lambda}{3}=-\frac{1}{\kappa^2},\qquad
\Phi(x)=-\frac12\Upsilon^2|x|^2,\qquad
\frac{d^2x}{dt^2}=-\operatorname{grad}\Phi=\Upsilon^2x .
\end{equation*}
Here $|x|$ is Euclidean distance from the origin on a real spatial slice. Negative $\Upsilon^2$ gives a restoring force and positive $\Upsilon^2$ a repulsive one. This homogeneous limit of the cosmological equations of motion is described by Gibbons and Patricot \cite{GP2003}. At the level of symmetries it is an In\"on\"u--Wigner contraction \cite{IW1953} of the de Sitter or anti-de Sitter algebra to one of the Newton--Hooke algebras in the classification of Bacry and L\'evy-Leblond \cite{BLL1968}. Dunajski realizes this contraction on cosmological twistor space \cite{D2023}.

Here the limit is taken directly in the field equations on spacetime. The construction begins with a holomorphic Cartan geometry; Lorentzian reality conditions are imposed in Section~\ref{IV}, after the field equations have been contracted over $\mathbb C$. In twinor theory \cite{Chien2026}, a massive field $Y$ of spin $s=\tfrac12,1,\tfrac32,\ldots$ has $2s$ symmetric spinor slots, each of which can carry either chirality. Its coupled covariant derivative $D$ acts on every slot and on any internal degrees of freedom. The starting equation and particle length are
\begin{equation*}
DY=J,\qquad \lambda=\frac{\hbar}{mc},
\end{equation*}
where $J$ is a one-form with values in the field bundle, $m$ is the fixed particle mass, and $\hbar$ is the reduced Planck constant. The choice of contracted components of $J$ fixes the mass coupling and any additional sources. Thus $\lambda$ decreases as $c^{-1}$ while $l$ grows as $c$; the mass term in the wave equation is fixed by the particle scale. In flat spacetime with trivial internal coupling, the free massive equations belong to the arbitrary-spin systems of Dirac and of Fierz and Pauli \cite{Dirac1936,FP1939}. With the mass normalization used here they take the symmetric multispinor form of Bargmann and Wigner, with one Dirac equation for each slot \cite{BW1948}. The reduction from the twinor equations is given in \cite[Sec.~III A]{Chien2026}.

After extracting the rest factor $e^{-imc^2t/\hbar}$, let $\psi$ denote the leading amplitude on the positive-rest-energy branch. In the homogeneous model, when the internal connection and the two source projections entering its evolution vanish, the projected equation is
\begin{equation*}
i\hbar\partial_t\psi=
\left(-\frac{\hbar^2}{2m}\Delta-\frac{m\Upsilon^2}{2}|x|^2\right)\psi ,
\qquad \Delta=\sum_{i=1}^3\frac{\partial^2}{\partial(x^i)^2}.
\end{equation*}
This equation follows by eliminating the first auxiliary spinor. The remaining slot equations still constrain its solutions. In particular, for $s\geq\tfrac32$ and $\Upsilon^2\ne0$, setting every limiting contracted source to zero forces $\psi=0$. The hierarchy is first derived on a static background with a general spatial metric and potential. Spin and internal curvature enter through the ordered kinetic term. The homogeneous model gives the displayed quadratic potential, and the scalar wave equation supplies the spin-zero reduction.

\section{Complex Twinor Geometry and Field Equations}\label{II}

Let $M$ be a complex four-manifold with a holomorphic reductive Cartan geometry of type $(Sp(4,\mathbb C),H)$ on the principal $H$-bundle $\mathcal P\to M$, where $H=SL(2,\mathbb C)_L\times SL(2,\mathbb C)_R$. Write $S$ and $S'$ for the rank-two bundles of the left and right defining representations. Nondegenerate soldering identifies $TM$ with $S\otimes S'$, the bundle associated to the reductive complement. The two factors are independent over $\mathbb C$. A prime labels the second factor; complex conjugation enters only on a real slice.

Unprimed spinor indices $A,B,\ldots$ range over $1,2$ and primed indices $A',B',\ldots$ over $1',2'$. Upper indices refer to the defining representations and lower indices to their duals. In $A_1,\ldots,A_{2s}$ the subscripts number the tensor slots. Spacetime indices range from $0$ to $3$, with $0$ temporal, and spatial indices $i,j,k$ from $1$ to $3$. Repeated contracted indices are summed. The alternating forms in the two spin spaces are normalized by
\begin{equation}
v_{A}=v^{B}\epsilon_{BA}, \qquad v^{A}=\epsilon^{AB}v_{B}, \qquad \epsilon_{12}=\epsilon^{12}=1, \qquad \epsilon^{AC}\epsilon_{BC}={\delta^{A}}_{B}.\label{eq:epsilon}
\end{equation}
The same convention holds in the primed space. Parentheses denote unit-weight symmetrization; indices between vertical bars are excluded. Choose a unit temporal direction and its adapted spin frame. Its normalized soldering identifies $S'^*$ with $S$, so a vector is represented by a $2\times2$ matrix. Write $\mathbf1$ for the identity, $\sigma_i$ for the Pauli matrices, and $\epsilon_{ijk}$ for the spatial alternating symbol. The temporal component is proportional to $\mathbf1$ and the spatial components are traceless. The complex bilinear spatial product $q$ and the normalization of the Pauli matrices are fixed by
\begin{gather}
v^{AA'}=\frac{1}{\sqrt{2}}(v^{0}\mathbf{1}+v^{i}\sigma_{i})^{AA'}, \qquad X=X^{i}\sigma_{i}, \qquad q(u,v)=\frac{1}{2}\operatorname{tr}(uv), \notag\\
\sigma_{i}\sigma_{j}=\delta_{ij}\mathbf{1}+i\epsilon_{ijk}\sigma_{k}, \qquad \sum_{i}{(\sigma_{i})^{B}}_{C}{(\sigma_{i})^{D}}_{E}=2{\delta^{B}}_{E}{\delta^{D}}_{C}-{\delta^{B}}_{C}{\delta^{D}}_{E}, \qquad \epsilon_{123}=1.\label{eq:pauli}
\end{gather}
Here $(T,X^1,X^2,X^3)$ are local complex coordinates and $X=X^i\sigma_i$ collects the spatial coordinates. Write $\theta$ for the temporal coframe one-form and $E=e^i\sigma_i$ for the spatial coframe. The vector fields $e_i$ are dual to $e^i$ on a spatial slice. Accordingly $\partial_E=\sum_i\sigma_i e_i$ differentiates in the coframe, which can differ from the coordinate frame. If $E\to E_0$, write $e_{0,i}$ for the limiting dual frame, $\partial_X=\sum_i\sigma_i e_{0,i}$ and $\nabla_{0,X}=\sum_i\sigma_i\nabla_{0,e_{0,i}}$ for its ordinary and covariant spinor derivatives. The homogeneous chart has $E_0=dX$.

\subsection{The Twinor Connection}

Let $c>0$, $l,m\in\mathbb{C}^{\times}$ and $\hbar\ne 0$, with
\begin{equation}
\Lambda=-3/l^{2}, \qquad \lambda=\hbar/(mc).\label{eq:scales}
\end{equation}
Write $\Gamma$ for the spatial rotation connection and $B$ for the rescaled boost one-form, both valued in traceless $2\times2$ matrices in the adapted frame. With the temporal and spatial coframes defined above, the Cartan connection in the fundamental representation and its metric are
\begin{equation}
A_{c}=\begin{pmatrix}\Gamma+B/(2c) & i(\theta\mathbf{1}+E)/(2l)\\ i(\theta\mathbf{1}-E)/(2l) & \Gamma-B/(2c)\end{pmatrix}, \qquad g_{c}=\theta\otimes\theta-q(E,E).\label{eq:geom}
\end{equation}
Here $q(E,E)(u,v)=q(E(u),E(v))$ for tangent vectors $u,v$. The diagonal blocks are the two Lorentz connections, and the off-diagonal blocks contain the soldering forms. Explicitly, $\omega_{L,R}=\Gamma\pm B/(2c)$, $\sqrt{2}e=\theta\mathbf{1}+E$ and $\sqrt{2}e^{\flat}=\theta\mathbf{1}-E$, where $e^{\flat}=\epsilon^{-1}e^{T}\epsilon$. The right connection $\omega_R$ acts on lower primed indices by the dual representation. The Lorentzian reality conditions are imposed in Section~\ref{IV}. Torsion-freeness is
\begin{equation}
d\theta+c^{-1}q(B\wedge E)=0, \qquad dE+[\Gamma,E]+c^{-1}B\wedge\theta=0.\label{eq:torsion}
\end{equation}
The brackets of forms are graded, and $q(B\wedge E)=\tfrac12\operatorname{tr}(B\wedge E)$. The homogeneous complexified de Sitter or anti-de Sitter model, denoted by (A)dS, is characterized locally by the full flatness condition $F_{A_c}=dA_c+A_c\wedge A_c=0$.

The fundamental twinor bundle, whose fibres are the direct sums of the two chiral carriers, is $\mathcal{T}=\mathcal{P}\times_{H}\mathbb{C}^{4}=S\oplus S'^{*}$. For $2s\in\mathbb{N}_{0}$, let $V$ be a complex associated vector bundle in a specified representation, with connection $a^{V}$, and define
\begin{equation}
E_{s}=\operatorname{Sym}^{2s}\mathcal{T}\otimes V, \qquad Y\in\Gamma(M,E_{s}), \qquad Y=\bigoplus_{k=0}^{2s}Y^{(2s-k,k)}.\label{eq:bundle}
\end{equation}
The spin ranges over $s=0,\tfrac12,1,\ldots$. The integer $k=0,\ldots,2s$ counts lower primed slots, and $p=2s-k$ counts upper unprimed slots. Each bidegree labels a rung. Its components are evaluations on fixed ordered slots, with no binomial factor. The component ${Y^{(2s-k,k)}{}_{A'_{1}\cdots A'_{k}}}^{A_{k+1}\cdots A_{2s}}$ lies in $\operatorname{Sym}^{2s-k}S\otimes\operatorname{Sym}^{k}S'^{*}\otimes V$ and is symmetric separately in its two spinor groups. With the adjacent mass normalization of \cite[Eqs.~(29), (37)--(40)]{Chien2026}, the fundamental particle connection and induced derivative are given by
\begin{equation}
A_{\lambda}=\begin{pmatrix}\omega_{L} & ie/(2\sqrt{2}\lambda)\\ ie^{\flat}/(2\sqrt{2}\lambda) & \omega_{R}\end{pmatrix}, \qquad D=d+A_{\lambda}^{(2s)}\otimes\mathbf{1}+\mathbf{1}\otimes a^{V}.\label{eq:particle}
\end{equation}
The particle connection $A_\lambda$ acts on each twinor slot. On the fundamental bundle, $A_\lambda=A_c$ precisely when $\lambda=l/2$. The derivative $\nabla^V$ contains the Lorentz and internal connections; the particle soldering terms are written separately.

\subsection{Twinor Field Equations}

The derivative index in $DY$ is a spinor pair. Contraction with an unprimed or primed field index gives the adjacent projections $\operatorname{ctr}=(\operatorname{ctr}_L,\operatorname{ctr}_R)$; contraction with both gives $\operatorname{ctr}_0$. The remaining projection $\operatorname{sym}$ is completely symmetric. Their index formulae are given in Appendix~\ref{A}. With $j_{\rm mass}[Y]$ the algebraic mass source and $j_{\rm phys},j_0$ the additional contracted sources, the prescription of \cite[Eq.~(39)]{Chien2026} is
\begin{equation}
\operatorname{ctr}J=j_{\text{mass}}[Y]+j_{\text{phys}}, \qquad \operatorname{ctr}_{0}J=j_{0}, \qquad \operatorname{sym}J=\operatorname{sym}(DY).\label{eq:sources}
\end{equation}
The last equality assigns the symmetric projection of the source without imposing an independent equation on the symmetric derivative. After subtraction of the algebraic mass source in Appendix~\ref{A}, the adjacent coefficient is $i/(\sqrt2\lambda)$ at every rung. The resulting adjacent and mixed equations are \cite[Eqs.~(28)--(29), (37)--(40)]{Chien2026}
\begin{align}
\operatorname{ctr}_{L}(\nabla^{V}Y^{(2s-k,k)})+\frac{imc}{\sqrt{2}\hbar}Y^{(2s-k-1,k+1)}&=j_{L,k}, & 0\le k<2s, \notag\\
\operatorname{ctr}_{R}(\nabla^{V}Y^{(2s-k-1,k+1)})+\frac{imc}{\sqrt{2}\hbar}Y^{(2s-k,k)}&=j_{R,k+1}, & 0\le k<2s,\label{eq:adjacent}
\\
\operatorname{ctr}_{0}(\nabla^{V}Y^{(2s-k,k)})&=j_{0,k}, & 1\le k<2s.\label{eq:mixed}
\end{align}
The source projections have the spinor symmetries displayed in \eqref{eq:adjacent}--\eqref{eq:mixed}. Applying $D$ to $DY=J$ relates the covariant derivative of the total one-form source to the curvature acting on the twinor \cite[Eqs.~(33)--(34), (67)--(68)]{Chien2026},
\begin{equation}
DJ=F_{D}Y, \qquad F_{D}=F_{\lambda}^{(2s)}\otimes\mathbf{1}+\mathbf{1}\otimes F_{a^{V}},\label{eq:compatible}
\end{equation}
where $F_{a^{V}}=da^{V}+a^{V}\wedge a^{V}$ and $F_\lambda$ includes both spin curvature and particle soldering terms. This identity is necessary for a prescribed total source. For a given interaction, the projected equations can impose further compatibility conditions on the fields and the background.

For $s=0$ there are no adjacent or mixed projections. The separate scalar equation is
\begin{equation}
(\Box_{g_{c}}^{V}+m^{2}c^{2}/\hbar^{2})Y=J_{\text{KG}}.\label{eq:scalar}
\end{equation}
Here $\Box_{g_c}^{V}$ is the metric trace of the covariant Hessian. Equation \eqref{eq:scalar} extends the flat scalar equation of \cite[Sec.~III A, following Eq.~(48)]{Chien2026} to a curved background with a source.

\section{The Newton--Hooke Limit}\label{III}

\subsection{Newton--Hooke Geometry}

To identify which symmetries survive the limit, resolve the Cartan connection into rotations $J(u)$, boosts $K_c(u)$, time translations $H_c$ and spatial translations $P_c(u)$. Here $J(u)$ denotes a generator, while the one-form $J$ in the field equation denotes the source. The generators associated with \eqref{eq:geom} are
\begin{equation}
J(u)=\begin{pmatrix}u & 0\\ 0 & u\end{pmatrix}, \qquad K_{c}(u)=\frac{1}{2c}\begin{pmatrix}u & 0\\ 0 & -u\end{pmatrix}, \qquad H_{c}=\frac{ic}{2l}\begin{pmatrix}0 & \mathbf{1}\\ \mathbf{1} & 0\end{pmatrix}, \qquad P_{c}(u)=\frac{i}{2l}\begin{pmatrix}0 & u\\ -u & 0\end{pmatrix},\label{eq:generators}
\end{equation}
where $u\in\mathfrak{sl}(2,\mathbb{C})$. Then $A_{c}=J(\Gamma)+K_{c}(B)+(\theta/c)H_{c}+P_{c}(E)$. Direct multiplication gives $[J(u),J(v)]=J([u,v])$, $[J(u),H_{c}]=0$, $[J(u),Z_{c}(v)]=Z_{c}([u,v])$ for $Z_{c}=K_{c},P_{c}$, and
\begin{gather}
[K_{c}(u),K_{c}(v)]=\frac{J([u,v])}{4c^{2}}, \qquad [P_{c}(u),P_{c}(v)]=\frac{J([u,v])}{4l^{2}}, \qquad [K_{c}(u),P_{c}(v)]=\frac{q(u,v)}{c^{2}}H_{c}, \notag\\
[K_{c}(u),H_{c}]=P_{c}(u), \qquad [H_{c},P_{c}(u)]=\frac{c^{2}}{l^{2}}K_{c}(u).\label{eq:brackets}
\end{gather}
At finite $c$ this is the complex (A)dS algebra in a rescaled basis.

Take $c\rightarrow\infty$ and $\Lambda\rightarrow0$ with $\kappa\in\mathbb{C}^{\times}$ fixed and
\begin{equation}
l=c\kappa\quad \Rightarrow \quad \Upsilon^{2}=\frac{c^{2}\Lambda}{3}=-\kappa^{-2},\label{eq:scaling}
\end{equation}
where the Newton--Hooke parameter $\Upsilon^2$ has dimensions of inverse time squared. It is the coefficient in the homogeneous geodesic equation $\ddot X=\Upsilon^2X$, with dots denoting $T$-derivatives. The contracted brackets are
\begin{equation}
[K,K]=[K,P]=[P,P]=0, \qquad [K(u),H]=P(u), \qquad [H,P(u)]=-\Upsilon^{2}K(u),\label{eq:nh}
\end{equation}
with the rotation brackets unchanged. The abstract contracted algebra is ten-dimensional. The limiting four-component representation has a nontrivial kernel.\footnote{Conjugate \eqref{eq:generators} by $S_c$ in \eqref{eq:transform} before taking the matrix limit. Then $J(w)=\mathrm{diag}(w,w)$, $H=i\,\mathrm{diag}(\mathbf1,-\mathbf1)/(2\kappa)$, and $K(u)+P(v)$ has only a lower-left block $(u+iv/\kappa)/2$. The kernel therefore consists precisely of $K(u)+P(v)$ with $u=-iv/\kappa$.}

The field calculation requires less symmetry than the homogeneous model. Assume the time slices are orthogonal to $\partial_T$ and the coframe is static. Let $N$ be the lapse, so $g_c(\partial_T,\partial_T)=c^2N^2$. Its first correction defines the potential $\Phi$. The subscript $T$ on a one-form denotes evaluation on $\partial_T$; the subscript $X$ denotes its spatial part. Assume a nondegenerate limiting spatial coframe $E_0$ and the following local expansions, with the same orders for the derivatives used below.
\begin{gather}
\theta=cN(X;c)dT, \qquad N=1+\Phi(X)/c^{2}+O(c^{-4}), \qquad E=E(X;c)=E_{0}(X)+O(c^{-2}), \notag\\
\Gamma_{T}=0, \qquad \Gamma_{X}=\Gamma_{0,X}(X)+O(c^{-2}), \qquad B_{X}=0, \qquad B_{T}=\beta_{c}=c^{2}\partial_{E}N=\beta+O(c^{-2}), \notag\\
\beta=\partial_{X}\Phi, \qquad a^{V}=\alpha+O(c^{-2}), \qquad \alpha=\alpha_{T}(T,X)dT+\alpha_{X}(T,X).\label{eq:static}
\end{gather}
Thus $\theta/c=dT+c^{-2}\Phi\,dT+O(c^{-4})$. With $h_c=q(E,E)$ and $h_0=q(E_0,E_0)$, the static metric gives
\begin{equation*}
\Gamma^i{}_{TT}=c^2Nh_c^{ij}\partial_jN\longrightarrow h_0^{ij}\partial_j\Phi,\qquad
\nabla^{h_0}_{\dot X}\dot X=-\operatorname{grad}_{h_0}\Phi ,
\end{equation*}
where dots denote $T$-derivatives. The limiting torsion-free connection preserves the clock $dT$ and the spatial inverse metric $h_0^{-1}$. On the Lorentzian slice this is the Newton--Cartan metric limit discussed in \cite[Thm.~3.1]{DG2016}, with parameter $c^{-2}$ and the opposite metric-sign convention. The time dependence allowed for the internal connection $\alpha$ in \eqref{eq:static} is a separate assumption on $V$.
\subsection{Field and Source Scaling}

In the adapted frame the temporal identification $\iota^{AA'}=\sqrt2 e^{AA'}(\partial_T)/(cN)$ has constant matrix entries. Its covariant derivative is $\nabla\iota=\omega_L\iota-\iota\omega_R=B/c$. The temporal boost coefficient is $\beta_c=B(\partial_T)$, with limit $\beta=\partial_X\Phi$. The time connection acts on the two carriers by $+\beta_c/(2c)$ and $-\beta_c/(2c)$, respectively.

For $2s>0$, the slot $a\in\{1,\ldots,2s\}$ in which the contraction is taken is called active; the remaining slots are spectators. Let $R_a=i\sqrt2\hbar cN(\tilde j_{L;a},\tilde j_{R;a})^T$ collect the raw sources of Appendix~\ref{A}. With $k$ primed spectators, $0\le k\le 2s-1$, the two entries come from rungs $k$ and $k+1$. Let $\tau_1,\tau_3$ be the Pauli matrices acting on chirality and $P_{c,a}=-i\hbar\sum_i\sigma_i^{(a)}\nabla^{\mathrm{rot},V}_{e_i}$ the spatial momentum. The derivative includes $\Gamma$ on every spinor slot and $a^V$ on $V$. In the adapted frame the inverse soldering matrices are $(\mathbf1,\sigma_i^T)/\sqrt2$ relative to $((cN)^{-1}\partial_T,e_i)$. Since $\epsilon\sigma_i^T\epsilon^T=-\sigma_i$, the two contractions have the same temporal derivative and opposite spatial derivatives. Multiplication by $i\sqrt2\hbar cN$ gives the exact active-slot equation
\begin{equation}
i\hbar\left(\partial_{T}+a_{T}^{V}+\sum_{b=1}^{2s}\frac{\tau_{3,b}\beta_{c}^{(b)}}{2c}\right)Y=cN\tau_{3,a}P_{c,a}Y+mc^{2}N\tau_{1,a}Y+R_{a}.\label{eq:hamilton}
\end{equation}
The rest operator $mc^2\tau_{1,a}$ has eigenvalue $mc^2$ on the selected branch for every active slot. After removal of the rest factor $e^{-imc^2T/\hbar}$, a component with $r$ minus slots is assumed to be $O(c^{-r})$ on compact subsets. Here $r=0,\ldots,2s$ counts minus eigenstates of $\tau_1$, whereas $k$ counts primed slots in the chirality basis. With $h_\pm$ the normalized row eigenvectors, the rescaling is
\begin{equation}
h_{\pm}=\frac{(1,\pm 1)}{\sqrt{2}}, \qquad S_{c}=\frac{1}{\sqrt{2}}\begin{pmatrix}\mathbf{1} & \mathbf{1}\\ c\mathbf{1} & -c\mathbf{1}\end{pmatrix}, \qquad \eta_{c}^{(r)}=e^{imc^{2}T/\hbar}c^{r}(h_{-}^{\otimes r}\otimes h_{+}^{\otimes(2s-r)})Y.\label{eq:transform}
\end{equation}

When the limits exist, write $\eta^{(r)}=\lim_{c\to\infty}\eta_c^{(r)}$, $\psi=\eta^{(0)}$, and $\chi=\eta^{(1)}$. The minus and plus indices of each component are separately symmetric. For $r$ minus spectators, the source limits are
\begin{align}
f_{a}^{(r)}&=\lim_{c\rightarrow\infty}e^{imc^{2}T/\hbar}c^{r}(h_{+}\otimes h_{-}^{\otimes r}\otimes h_{+}^{\otimes(2s-r-1)})R_{a}, \notag\\
g_{a}^{(r)}&=\lim_{c\rightarrow\infty}e^{imc^{2}T/\hbar}c^{r-1}(h_{-}\otimes h_{-}^{\otimes r}\otimes h_{+}^{\otimes(2s-r-1)})R_{a}, \qquad 0\le r<2s.\label{eq:sourcelimits}
\end{align}

The first factor acts on slot $a$. The power $c^{r-1}$ includes division of the transformed minus row by $c^2$. The two limits come from the same finite-$c$ source, whose raw components satisfy the trace relations in Appendix~\ref{A}.

\subsection{Coupled Nonrelativistic Field Equations}

Let $\rho(\Gamma_0)$ denote the sum of the limiting rotation actions on the spinor indices. The limiting internal connection $\alpha$ acts once on $V$. Define the coupled derivative, spatial momentum and time evolution operator by
\begin{equation}
\nabla_{0}=d+\rho(\Gamma_{0})+\alpha, \qquad {P^{A}}_{B}=-i\hbar{\nabla_{0,X}{}^{A}}_{B}, \qquad \mathcal{H}=i\hbar\nabla_{0,T}-m\Phi, \qquad \nabla_{0,T}=\partial_{T}+\alpha_{T}.\label{eq:operators}
\end{equation}
Write $P_a=-i\hbar\sum_i\sigma_i^{(a)}\nabla_{0,e_{0,i}}$. In the hierarchy below the spinor indices are suppressed and all slots are kept in their original positions. In a momentum term the active slot changes sign between the input and output. In a boost term only the indicated spectator changes sign.
\begin{theorem}
Let $l=c\kappa$ with fixed $\kappa\ne0$. Fix $2s\in\mathbb N_{>0}$ and an exact family of solutions of \eqref{eq:adjacent}--\eqref{eq:mixed} on a complex background satisfying \eqref{eq:static}, with fixed nonzero $m$ and $\hbar$. Suppose the rescaled components \eqref{eq:transform} and their first derivatives converge locally uniformly to $\eta^{(r)}$ and its first derivatives, and the coupled sources have the limits \eqref{eq:sourcelimits}. For each active slot $a$ and each choice of $r$ minus spectators, $0\le r<2s$, the limiting fields satisfy
\begin{align}
\mathcal H\eta^{(r)}&=P_a\eta^{(r+1)}-\frac{i\hbar}{2}\sum_{b\ {\rm minus\ spectator}}\beta^{(b)}\eta^{(r-1)}+f_a^{(r)},\label{eq:evolution}\\
2m\eta^{(r+1)}&=P_a\eta^{(r)}+g_a^{(r)}.\label{eq:constraint}
\end{align}
In the first row the active output slot is plus. For each of its $r$ minus spectators $b$, the input to $\beta^{(b)}$ has slot $b$ plus. The sum is empty at $r=0$. In the second row the active output slot is minus; at $r=2s-1$ this gives the all-minus component. The equations hold before symmetrization over the active slot.
\end{theorem}

\begin{proof}
Conjugate \eqref{eq:hamilton} by $S_c$ on every slot and remove the common rest factor. On the active slot, $S_c\tau_1S_c^{-1}=\tau_3$, while $cS_c\tau_3S_c^{-1}$ has off-diagonal entries $1,c^2$ in the upper-right and lower-left blocks, respectively. The plus row therefore retains $P_a\eta^{(r+1)}$ and the mass correction $mc^2(N-1)\to m\Phi$. Each minus spectator contributes the lower-left boost $\beta_c/2\to\beta/2$ on the left-hand side; the upper-right boost is $O(c^{-2})$. This gives \eqref{eq:evolution}. After the minus row is divided by $c^2$, its time derivative and boosts vanish because the rescaled fields and their first derivatives are locally uniformly bounded, its momentum term tends to $P_a\eta^{(r)}$, and its mass coefficient $-m(N+1)$ tends to $-2m$. This gives \eqref{eq:constraint}. The respective source weights are $c^r$ and $c^{r+1}/c^2=c^{r-1}$, as in \eqref{eq:sourcelimits}.
\end{proof}

At $r=0$, with the minus slot of $\chi$ placed at $a$, the two rows read
\begin{equation*}
\mathcal H\psi=P_a\chi+f_a^{(0)},\qquad 2m\chi=P_a\psi+g_a^{(0)}.
\end{equation*}
Since $P_a\psi+g_a^{(0)}=2m\chi$ is continuously differentiable, its derivative can be taken pointwise. Separating the two terms distributionally and averaging over the active slot gives
\begin{equation}
(i\hbar\nabla_{0,T}-m\Phi)\psi^{A_{1}\cdots A_{2s}}=\frac{1}{2m}{P^{(A_{1}}}_{B}{P^{B}}_{C}\psi^{|C|A_{2}\cdots A_{2s})}+f^{(0)(A_{1}\cdots A_{2s})}+\frac{1}{2m}{P^{(A_{1}}}_{B}g^{(0)|B|;A_{2}\cdots A_{2s})}.\label{eq:reducedagain}
\end{equation}
The expanded equation is classical if $\psi$ has two continuous spatial derivatives and $g_a^{(0)}$ has one. Local uniform convergence of holomorphic rescaled fields on a common complex neighborhood also yields their spatial derivatives by the Cauchy formula, in a holomorphic frame with holomorphic coefficients. For $2s>1$, each slot equation has the same left-hand side $\mathcal H\psi$. Hence the differences between its right-hand sides for two choices of $a$ must vanish. Averaging discards these differences. Equation~\eqref{eq:reducedagain} is a necessary equation for a regular limit; the remaining rows of the hierarchy must still be satisfied.

\subsection{The Homogeneous Limit}

For the homogeneous complexified model, choose the branch $N\rightarrow 1$ with
\begin{gather}
N^{2}=1-\frac{\Upsilon^{2}q(X,X)}{c^{2}}, \qquad \theta=cNdT, \qquad E=dX+\frac{\Upsilon^{2}Xq(X,dX)}{c^{2}N(1+N)}, \notag\\
B=-\Upsilon^{2}X\,dT, \qquad \Gamma=\frac{\Upsilon^{2}[X,dX]}{4c^{2}(1+N)}.\label{eq:homogeneous}
\end{gather}
On fixed domains with $N(1+N)\ne 0$,
\begin{equation}
E_{0}=dX, \qquad \Gamma_{0}=0, \qquad \Phi=-\frac{1}{2}\Upsilon^{2}q(X,X), \qquad B_{0}=(\partial_{X}\Phi)dT.\label{eq:model}
\end{equation}

To check Cartan flatness, put $K=\Upsilon^2/c^2=-1/l^2$ and let $D^\Gamma$ be the spatial exterior covariant derivative in the adjoint representation. Using $dN=-Kq(X,dX)/N$ and $[[X,u],X]=4(Xq(X,u)-q(X,X)u)$ in \eqref{eq:homogeneous} gives
\begin{equation*}
q(X,E)=\frac{q(X,dX)}{N},\qquad D^\Gamma E=0,\qquad D^\Gamma X=NE,\qquad
F_\Gamma(u,v)=\frac K4[E(u),E(v)]
\end{equation*}
for spatial tangent vectors $u,v$ in the last identity. The first two identities give the two torsion equations. The last cancels the spatial translation bracket in \eqref{eq:brackets}; $D^\Gamma X=NE$ cancels the mixed boost curvature. Thus $F_{A_c}=0$. The limiting affine connection has $\Gamma^i{}_{TT}=-\Upsilon^2 X^i$, giving $\ddot X=\Upsilon^2X$.

Take $\alpha=0$ and suppose $f_a^{(0)}=g_a^{(0)}=0$ for each active slot. The spatial momenta now have constant Pauli matrices and commuting derivatives, so $p_a^2=-\hbar^2\Delta_X$, where $\Delta_X=\sum_i\partial^2/\partial(X^i)^2$. The projected evolution equation and the first constraint reduce to
\begin{equation}
i\hbar\partial_{T}\psi=\left(-\frac{\hbar^{2}}{2m}\Delta_{X}-\frac{m\Upsilon^{2}}{2}q(X,X)\right)\psi, \qquad 2m\chi^{B;A_{2}\cdots A_{2s}}={p^{B}}_{C}\psi^{CA_{2}\cdots A_{2s}}, \qquad p=-i\hbar\partial_{X}.\label{eq:sch}
\end{equation}
For $s=0$ the same homogeneous evolution follows from the source-free scalar equation \eqref{eq:scalar} with $\alpha=0$. After removal of the rest factor, the surviving terms are $-2im\partial_T\psi/\hbar-\Delta_X\psi+2m^2\Phi\psi/\hbar^2=0$, provided the amplitudes and their first time and first two spatial derivatives converge locally uniformly and their second time derivatives remain locally uniformly bounded.

For $2s\geq3$, choose an active slot $a$ and two minus spectators $b,d$. Suppose $\alpha=0$, all $g_a^{(r)}=0$, and $f_a^{(0)}=f_a^{(1)}=0$. With $p_b=-i\hbar\sigma_i^{(b)}\partial_i$, the constraints give $(2m)^2\eta^{(2)}=p_bp_d\psi$. In the $r=2$ evolution equation, the terms in $[\mathcal H,p_bp_d]$ containing first derivatives of $\Phi$ cancel the two spectator boosts. The Hessian term gives
\begin{equation*}
f_a^{(2)}=-\frac{\hbar^2}{4m}\sigma_i^{(b)}\sigma_j^{(d)}(\partial_i\partial_j\Phi)\psi
=\frac{\hbar^2\Upsilon^2}{4m}\psi,
\end{equation*}
where $\partial_i\partial_j\Phi=-\Upsilon^2\delta_{ij}$ and $\sum_i\sigma_i^{(b)}\sigma_i^{(d)}\psi=\psi$ by \eqref{eq:pauli} and the symmetry of $\psi$. Thus, when $\Upsilon^2\ne0$, vanishing of all limiting sources forces $\psi=0$ and hence, by \eqref{eq:constraint}, all limiting auxiliaries vanish.

\section{Lorentzian Real Forms}\label{IV}

On a Lorentzian real slice, take $T=t\in\mathbb R$, $X=x=x^\dagger$ traceless, and $m,\hbar>0$. In an adapted spin frame the spatial spin group is $SU(2)$, $q(x,x)=|x|^2$, and dualization of the primed carrier gives
\begin{equation}
\bar\theta=\theta,\qquad E^\dagger=E,\qquad \omega_R=-\omega_L^\dagger,\qquad \Gamma^\dagger=-\Gamma,\qquad B^\dagger=B.\label{eq:real}
\end{equation}
The particle length $\lambda=\hbar/(mc)$ is positive on both real forms. In the homogeneous model the choice of cosmological scale determines the sign of the potential in \eqref{eq:sch}. For $\kappa_0>0$,
\begin{align}
\mathrm{AdS}:&\quad \kappa=\kappa_0, & \Upsilon^2&=-\kappa_0^{-2}, & m\Phi&=\frac{m|x|^2}{2\kappa_0^2},\notag\\
\mathrm{dS}:&\quad \kappa=-i\kappa_0, & \Upsilon^2&=\kappa_0^{-2}, & m\Phi&=-\frac{m|x|^2}{2\kappa_0^2}.\label{eq:signs}
\end{align}
These are the oscillator and inverted-oscillator potentials, respectively. For positive spin, internal coupling is included through the connection $\alpha$ in \eqref{eq:operators}, and forcing through the projected sources in \eqref{eq:reducedagain}. For $s=\tfrac12$ with trivial internal coupling and vanishing projected sources, the flat limit $\Upsilon^2\to0$ gives the free L\'evy-Leblond system \cite[Eqs.~(25), (34)]{LL1967}, with $\chi_{\rm LL}=-\chi$ in units $\hbar=1$.

\appendix

\section{Spinor Contractions and Source Projections}\label{A}

Let $p=2s-k$. After inverse soldering, the one-form index is the pair $BB'$. The remaining indices belong to the field at rung $k$. For a one-form coefficient ${\Xi_{BB'A'_{1}\cdots A'_{k}}}^{A_{k+1}\cdots A_{2s}}$, suppress the rung superscript and define, as in \cite[Eq.~(28)]{Chien2026},
\begin{align}
{(\operatorname{ctr}_{L}\Xi)_{B'A'_{1}\cdots A'_{k}}}^{A_{k+2}\cdots A_{2s}}&={\Xi_{B(B'A'_{1}\cdots A'_{k})}}^{BA_{k+2}\cdots A_{2s}}, \notag\\
{(\operatorname{ctr}_{R}\Xi)_{A'_{2}\cdots A'_{k}}}^{CA_{k+1}\cdots A_{2s}}&=\epsilon^{(C|B|}\epsilon^{D'B'}{\Xi_{BB'D'A'_{2}\cdots A'_{k}}}^{A_{k+1}\cdots A_{2s})}, \notag\\
{(\operatorname{ctr}_{0}\Xi)_{A'_{2}\cdots A'_{k}}}^{A_{k+2}\cdots A_{2s}}&=\epsilon^{D'B'}{\Xi_{BB'D'A'_{2}\cdots A'_{k}}}^{BA_{k+2}\cdots A_{2s}}, \notag\\
{(\operatorname{sym}\Xi)_{B'A'_{1}\cdots A'_{k}}}^{CA_{k+1}\cdots A_{2s}}&=\epsilon^{(C|B|}{\Xi_{B(B'A'_{1}\cdots A'_{k})}}^{A_{k+1}\cdots A_{2s})}.\label{eq:projections}
\end{align}

Symmetrization and trace in the two spin spaces give the four projections in \eqref{eq:projections}, with absent-index terms omitted. In the left contraction the active slot contributes twice and the $2s-k-1$ unprimed spectators contribute once each. The left and right soldering coefficients at rung $k$ are therefore $i(2s-k+1)/(2\sqrt2\lambda)$ and $i(k+1)/(2\sqrt2\lambda)$, respectively. Soldering changes the rung by one. Its two neighboring bidegrees $(2s-k\mp1,k\pm1)$ differ from both the mixed and the fully symmetric target bidegrees, so neither projection receives a soldering term. The mass subtraction is
\begin{equation}
(j_{\text{mass}})_{L,k}=\frac{i(2s-k-1)}{2\sqrt{2}\lambda}Y^{(2s-k-1,k+1)}, \qquad (j_{\text{mass}})_{R,k}=\frac{i(k-1)}{2\sqrt{2}\lambda}Y^{(2s-k+1,k-1)}.\label{eq:mass}
\end{equation}
Subtraction leaves $i/(\sqrt2\lambda)$ at each adjacent rung, with $0\le k<2s$ on the left and $1\le k\le2s$ on the right. The raw sources retain the derivative index before symmetrization with the spectators.
\begin{align}
{(\tilde{j}_{L,k})_{B'A'_{1}\cdots A'_{k}}}^{A_{k+2}\cdots A_{2s}}&={(j_{L,k})_{B'A'_{1}\cdots A'_{k}}}^{A_{k+2}\cdots A_{2s}}-\frac{k}{k+1}\epsilon_{B'(A'_{1}}{(j_{0,k})_{A'_{2}\cdots A'_{k})}}^{A_{k+2}\cdots A_{2s}}, \notag\\
{(\tilde{j}_{R,k})_{A'_{2}\cdots A'_{k}}}^{CA_{k+1}\cdots A_{2s}}&={(j_{R,k})_{A'_{2}\cdots A'_{k}}}^{CA_{k+1}\cdots A_{2s}}+\frac{2s-k}{2s-k+1}\epsilon^{C(A_{k+1}}{(j_{0,k})_{A'_{2}\cdots A'_{k}}}^{A_{k+2}\cdots A_{2s})}.\label{eq:raw}
\end{align}
The left raw row occurs for $p=2s-k>0$ and the right raw row for $k>0$. The lower trace uses $\epsilon^{D'B'}$ on the first spectator and derivative indices, respectively, and the upper trace uses $\epsilon_{CA_1}$ on the derivative and first spectator. In the normalized symmetrization, the direct lower contraction contributes $-2$ and each of the $k-1$ exchanges contributes $-1$. Thus the lower trace of $\epsilon_{B'(A'_1}M_{A'_2\cdots A'_k)}$ is $-(k+1)M/k$ for $k>0$. The corresponding upper contributions are $+2$ and $p-1$ copies of $+1$. The upper trace of $\epsilon^{C(A_1}M^{A_2\cdots A_p)}$ is $(p+1)M/p$ for $p=2s-k>0$. These traces fix the coefficients in \eqref{eq:raw}, with the corresponding terms omitted at the endpoints.

\begin{acknowledgments}
I thank Maciej Dunajski for suggesting the Newton–Hooke limit as a direction for this work during the CaLIGOLA Dissemination and Conclusion Workshop in Valtice.

I was supported by the Czech Science Foundation (GA\v{C}R) under the research project ``Cartan supergeometries and Higher Cartan geometries'' (Grant No.~GA24-10887S) and by the European Union's Horizon Europe research and innovation programme under the Marie Sk\l{}odowska-Curie grant agreement No.~101086123 (CaLIGOLA). This research was supported by the Simons Foundation grant SFI-MPS-T-Institutes-00010825 and by State Treasury funds as part of a task commissioned by the Minister of Science and Higher Education under the project ``Organization of the Simons Semesters at the Banach Center --- New Energies in 2026--2028'' (MNiSW/2025/DAP/491).
\end{acknowledgments}


\begin{thebibliography}{10}

\bibitem{GP2003}
G. W. Gibbons and C. E. Patricot, \textit{Newton--Hooke spacetimes, Hpp-waves and the cosmological constant}, Class. Quantum Grav. \textbf{20} (2003), 5225--5239, \href{https://doi.org/10.1088/0264-9381/20/23/016}{doi:10.1088/0264-9381/20/23/016}, \href{https://arxiv.org/abs/hep-th/0308200}{arXiv:hep-th/0308200}.

\bibitem{IW1953}
E. In\"on\"u and E. P. Wigner, \textit{On the contraction of groups and their representations}, Proc. Natl. Acad. Sci. USA \textbf{39} (1953), 510--524, \href{https://doi.org/10.1073/pnas.39.6.510}{doi:10.1073/pnas.39.6.510}.

\bibitem{BLL1968}
H. Bacry and J.-M. L\'evy-Leblond, \textit{Possible kinematics}, J. Math. Phys. \textbf{9} (1968), 1605--1614, \href{https://doi.org/10.1063/1.1664490}{doi:10.1063/1.1664490}.

\bibitem{D2023}
M. Dunajski, \textit{Equivalence principle, de-Sitter space, and cosmological twistors}, Int. J. Mod. Phys. D \textbf{32}, no.~14 (2023), 2341001, \href{https://doi.org/10.1142/S0218271823410018}{doi:10.1142/S0218271823410018}, \href{https://arxiv.org/abs/2304.08574}{arXiv:2304.08574} [gr-qc].

\bibitem{Chien2026}
C.-H. Chien, \textit{Twinor Theory: Field Equations from $(A)dS_4^{\mathbb C}$ Spin-Tractors: A Twin to Twistor Theory Modelled on $(A)dS_4^{\mathbb C}$}, \href{https://arxiv.org/abs/2610.40435v1}{arXiv:2610.40435v1} [math-ph], 29 October 2026.

\bibitem{Dirac1936}
P. A. M. Dirac, \textit{Relativistic wave equations}, Proc. R. Soc. Lond. A \textbf{155} (1936), 447--459, \href{https://doi.org/10.1098/rspa.1936.0111}{doi:10.1098/rspa.1936.0111}.

\bibitem{FP1939}
M. Fierz and W. Pauli, \textit{On relativistic wave equations for particles of arbitrary spin in an electromagnetic field}, Proc. R. Soc. Lond. A \textbf{173} (1939), 211--232, \href{https://doi.org/10.1098/rspa.1939.0140}{doi:10.1098/rspa.1939.0140}.

\bibitem{BW1948}
V. Bargmann and E. P. Wigner, \textit{Group theoretical discussion of relativistic wave equations}, Proc. Natl. Acad. Sci. USA \textbf{34} (1948), 211--223, \href{https://doi.org/10.1073/pnas.34.5.211}{doi:10.1073/pnas.34.5.211}.

\bibitem{DG2016}
M. Dunajski and J. Gundry, \textit{Non-relativistic twistor theory and Newton--Cartan geometry}, Commun. Math. Phys. \textbf{342} (2016), 1043--1074, \href{https://doi.org/10.1007/s00220-015-2557-8}{doi:10.1007/s00220-015-2557-8}, \href{https://arxiv.org/abs/1502.03034}{arXiv:1502.03034}.

\bibitem{LL1967}
J.-M. L\'evy-Leblond, \textit{Nonrelativistic particles and wave equations}, Commun. Math. Phys. \textbf{6} (1967), 286--311, \href{https://doi.org/10.1007/BF01646020}{doi:10.1007/BF01646020}.

\end{thebibliography}
\end{document}